# STUDY OF NIOBIUM SURFACE UNDER ULTRA HIGH VACUUM AFTER HEAT TREATMENTS FOR SRF CAVITIES*

C. Boutelaa†,[1,2], B. Laxdal[2], P. Kolb[2], C. Craig[2,3], B. Mercier[1], G. Sattonnay[1], D. Longuevergne[1]
[1]IJCLab, Paris Saclay University, Orsay, France
[2]TRIUMF, Vancouver, Canada
[3]Western University, Ontario, Canada

*Abstract*

Specific heat treatments applied to superconducting radio-frequency (SRF) cavities, such as nitrogen infusion or Mid-T baking, aim to improve the quality factor ($Q_0$) at medium accelerating fields (˜10‑20 MV/m). These treatments reduce the BCS surface resistance by tuning the mean free path of niobium over a few hundred nanometers, either by diffusing oxygen from the native oxide layer or by diffusing nitrogen after the dissolution of the oxide layer. However, these treatments preclude the usual chemical polishing, as it would reverse the beneficial effects of the heat treatments, making the cavities highly sensitive to surface contamination. In particular, the formation of niobium carbides, which can mask the expected benefits, strongly depends on the annealing conditions, surface preparation, and the material's history. Several hypotheses are considered regarding the origin of carbon: vacuum contamination, surface pollution, or internal migration from the niobium itself, potentially enriched with carbon during previous chemical treatments (BCP, EP). This work aims to identify the primary source of carbon responsible for niobium carbide growth, using techniques such as X-ray photoelectron spectroscopy (XPS), scanning electron microscopy (SEM), and secondary ion mass spectrometry (SIMS). The study will also help pinpoint the key influencing parameters, thereby contributing to a better understanding of and potential mitigation strategies for their impact on SRF cavity performance.

## INTRODUCTION

The superiority of superconducting cavities over normal-conducting metallic cavities lies in their ability to store large amounts of energy while dissipating significantly less power. The performance of SRF (Superconducting Radio Frequency) cavities is evaluated in terms of the quality factor $Q_0$ = G/Rs, where G is the geometric factor, which depends on the cavity geometry, and Rs is the surface resistance, which varies with the accelerating gradient Eacc. A high-quality factor can be achieved by reducing the surface resistance of SRF cavities. This resistance consists of two main components: the residual resistance ($R_{RES}$), which is temperature-independent, and the BCS resistance ($R_{BCS}$), which is temperature-dependent [1, 2]. Sources of $R_{RES}$ include trapped magnetic flux during cooldown, impurities, hydrides and oxides, surface imperfections, and contamination. The surface resistance and its behaviour versus accelerating gradient can be tuned depending on the requirements. This tuning is in part performed by applying specific heat treatments ranging from 100 °C up to 800 °C. During cavity heat treatments, carbon can lead to the precipitation of niobium carbides [3, 4], Carbon can originate from the bulk, surface contamination or residual gas. The carbides are believed to be responsible of Q-factor degradation (Fig. 1) subsequent to new medium temperature heat treatments (Mid-T baking) [5]. This surface carburation is known since the 80s [6], but carbide formation was not a problem as a flash BCP or EP treatment was systematically performed after heat treatment removing thus any precipitates. This is not the case anymore with new surface preparation Mid-T baking. In an effort to understand what are the main sources of carbon and what are the main parameters driving the carbide formation, we initiated a sample study. A standard procedure to prepare niobium samples was established. Niobium sheets with an RRR of 300 were cut using water jet cutting at TRIUMF to perform multiple high-vacuum heat treatments at temperatures ranging from 100 °C to 700 °C in 50 °C increments (under a vacuum of ~$10^{-8}$ Torr), as well as a treatment at 800 °C under $10^{-7}$ Torr, each lasting 3 hours. In this article, we present several high-vacuum heat treatments performed using TRIUMF's induction furnace, along with a surface study of the thermally treated samples using X-ray photoelectron spectroscopy (XPS) at IJCLab and secondary ion mass spectrometry (SIMS) in Western University. In the last part, we will show how the heating environment can influence the formation of precipitates through heat treatments under UHV, performed in the Supratech furnace at IJCLab.

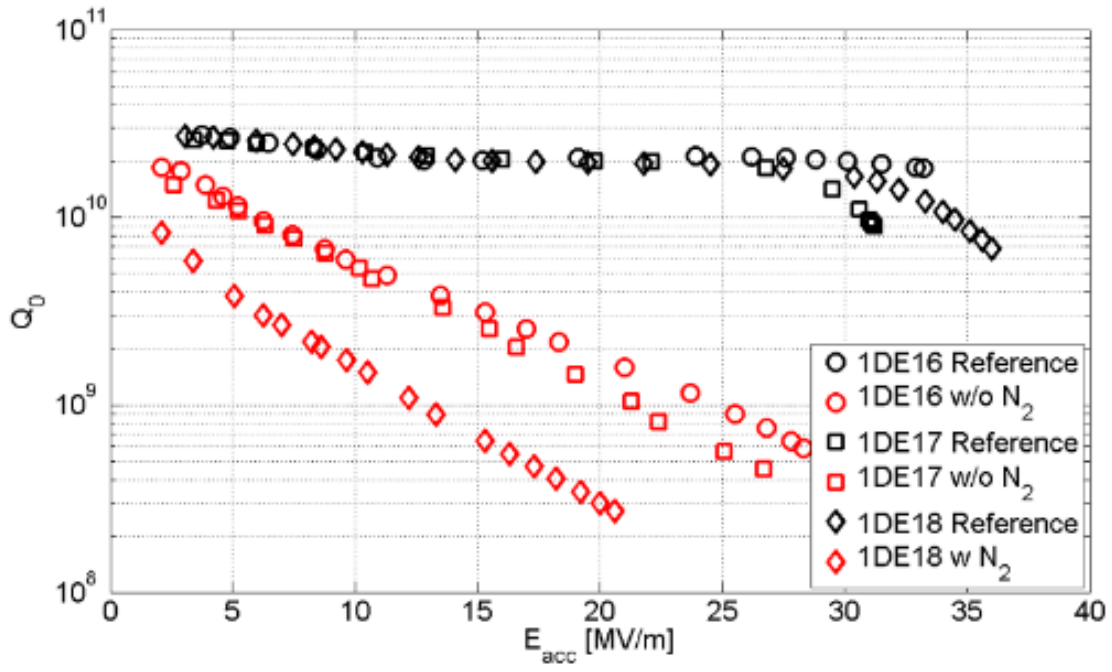

Figure 1: Quality factor $Q_0$ versus accelerating field Eacc of the three single-cell cavities taken at 2 k [4].

* Work supported by the CNRS-TRIUMF International Research Laboratory NPAT (Nuclear Physics, Nuclear Astrophysics and Accelerator Technology)
† chahinez.boutelaa@ijclab.in2p3.fr

## FABRICATION OF Nb SAMPLES

Niobium was cut by water jet into 10 mm × 10 mm squares from fine-grain niobium manufactured by Tokyo. Each sample was chemically treated in a buffered chemical solution (HF: $HNO_3$:$H_3PO_4$ = 1:1:2 by volume, removing ~100 µm) at room temperature, followed by thorough rinsing with ultra-pure water. Thermal treatments of the niobium samples were carried out in the TRIUMF induction furnace at 800 °C for 3 hours. Afterwards, each sample was subjected to a BCP flash etching of about 10 µm. The samples were not exposed to a final High-Pressure Rinsing (HPR) as would a real cavity. The results obtained might thus not reflect exactly what would happen in a cavity. The goal was here to study the worst-case scenario.

## TRIUMF FURNACE: MID -T BAKING

The SRF group at TRIUMF has a UHV furnace used for the heat treatment of cavities and samples (Fig. 2). The furnace consists of an RF generator, an induction coil, and a niobium susceptor. Details regarding the vacuum and cooling systems will not be discussed here. The temperature inside the chamber is not controlled directly; instead, the AC current supplied by the RF module to the coil is adjusted. This current generates an oscillating magnetic field, which is coupled into the conductive niobium susceptor wrapped by the coil. The field induces currents within the susceptor, which heats up due to its electrical resistance. The heat is then transferred by radiation to objects placed inside the susceptor, such as a cavity or samples. The temperature inside the chamber is measured using a thermocouple.

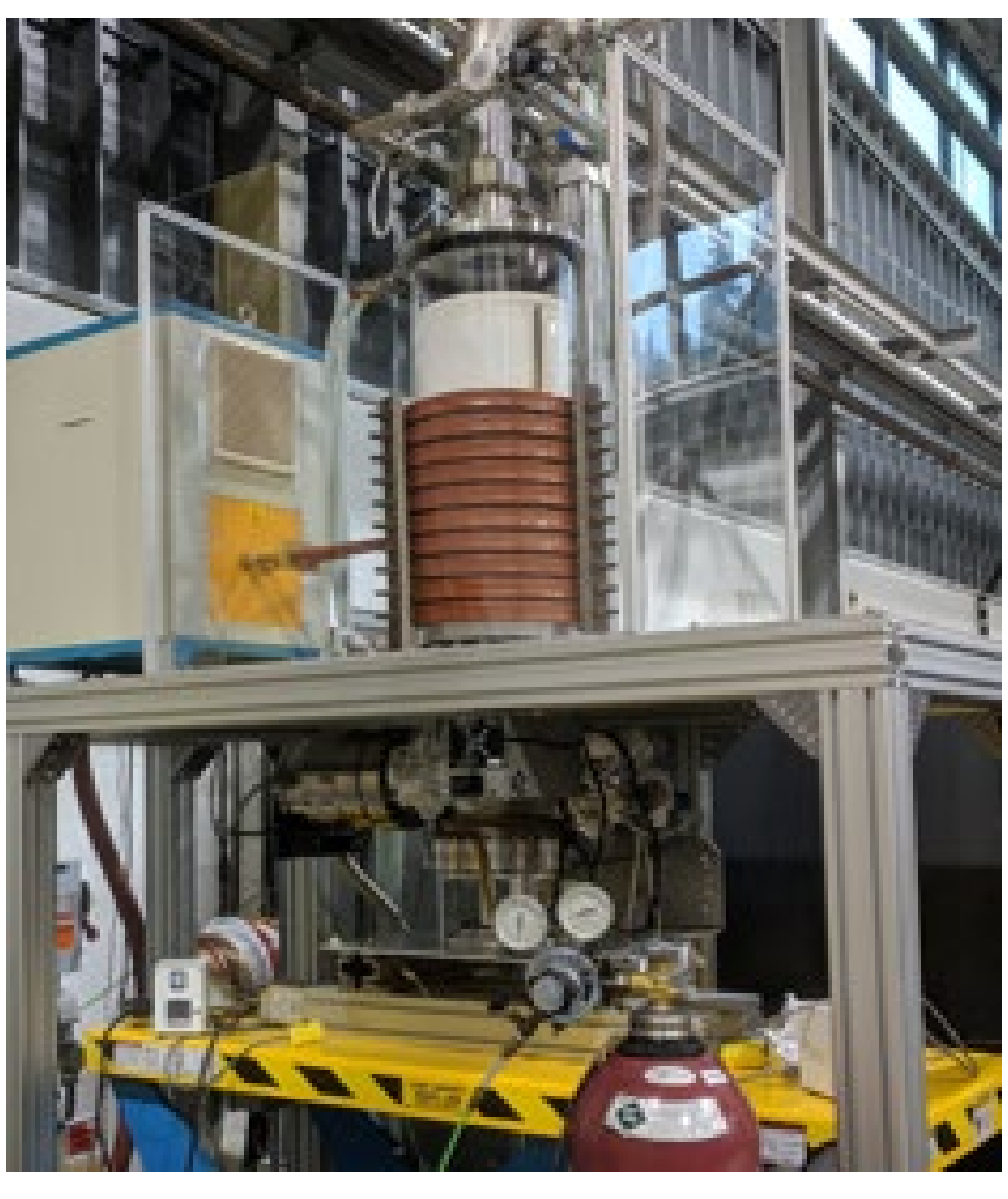

Figure 2: TRIUMF induction furnace.

The heat treatment was performed in the range form 100 °C to 800 °C during 3 hrs and base pressure $1.10^{-9}$ Torr (see the Fig. 3).

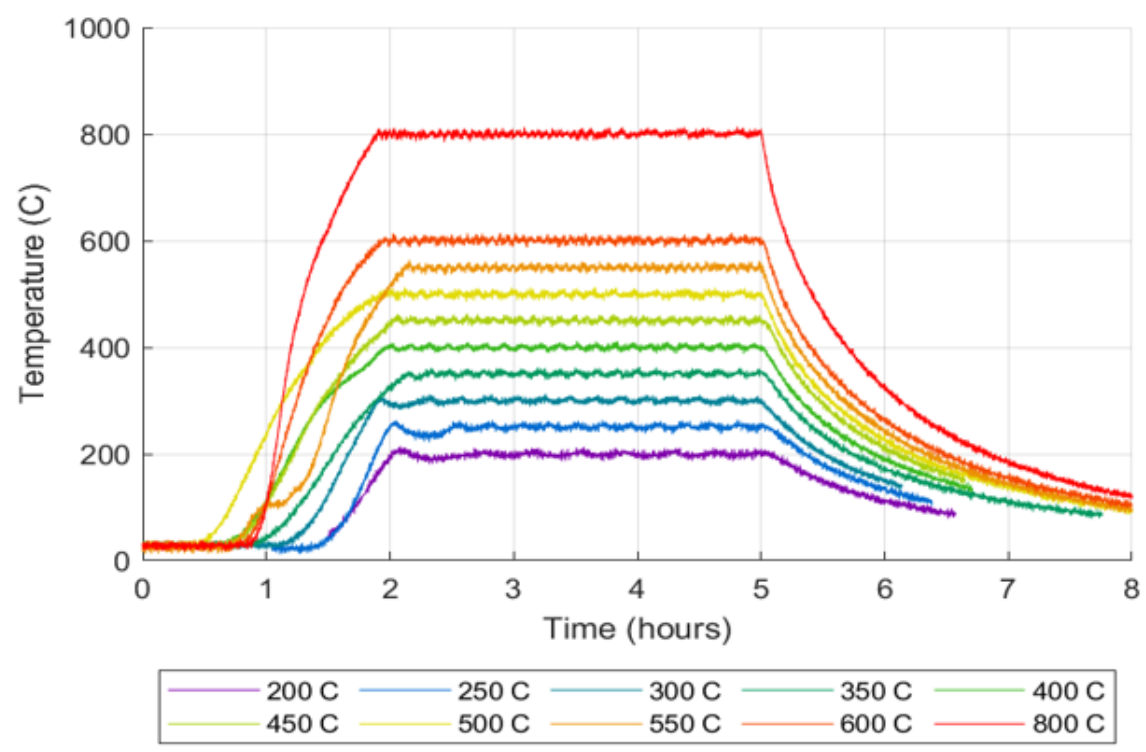

Figure 3: Temperature vs Time during mid-T baking

The heat treatment procedure consists of three stages: heating-up, baking under UHV ($10^{-7}$ to $10^{-8}$ Torr), and cooling-down. Once baking was completed, the sample heating was switched off and the sample was allowed to cool naturally, mainly by radiation. The residual gas analyzer used to monitor gas species in the analysis chamber during the heat treatment was scanning the following molecules: CO and $CO_2$ (see Fig. 4).

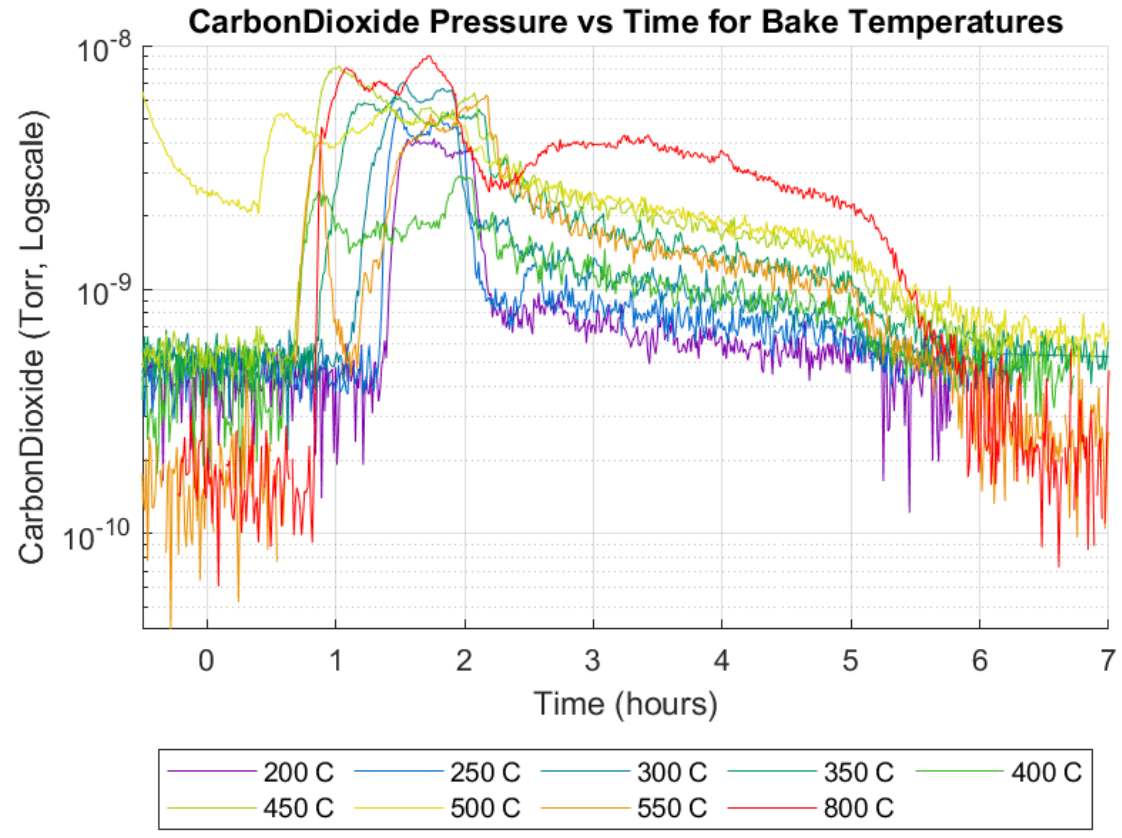

Figure 4: Carbon dioxide pressure vs Time during mid-T baking.

## STUDY OF THE CARBIDE FORMATION

The surface of the Nb samples treated in the TRIUMF induction furnace was characterized primarily by X-ray photoemission spectroscopy (XPS). This technique enabled the determination of binding energies of occupied states in the core-level region, allowing the identification of elemental chemical states as well as an estimation of the niobium carbide concentration at the surface. All XPS measurements were carried out after the baking process (see Fig. 5).

Figure 5: Evolution of niobium carbides from 25 °C to 800 °C as analyzed by XPS.

A number of samples were analyzed at different sputtering levels ranging from 0 to 60 seconds using an energy of 200 eV. All spectra were fitted using the Avantage software package. The XPS spectra were collected from at least three randomly selected locations on the sample surface, both before (at room temperature) and after heat treatment, to evaluate the uniformity of carbide formation. The C 1s core level is associated with carbonaceous adsorbates a contamination layer commonly present on solid surfaces after air exposure for each spectrum. The Fig. 6, is an example for 350 °C.

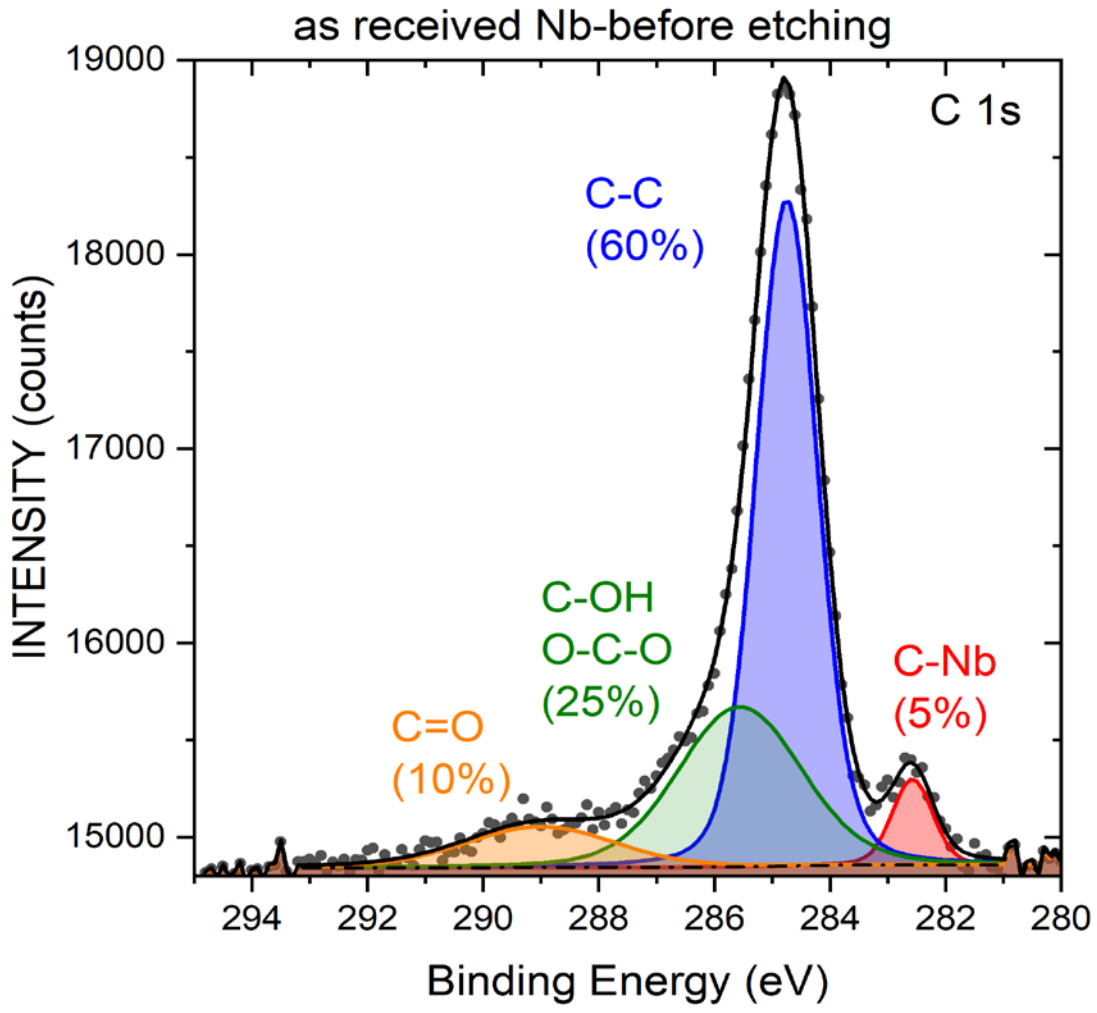

Figure 6: XPS fitting for 350 °C for 3 hrs.

This peak was fitted with four components. The most intense peak appears at 285.1 eV. The lower intensity peaks are attributed to O–C–O (285.7 eV), C=O (289.2 eV). The Nb–C peak is observed at 282.6 eV. Figure 5 shows the quantification of niobium carbides (Nb–C) as a function of temperature. At temperatures below 250 °C, no carbide formation was observed, as the oxide dissolution process was not yet complete. Carbide formation begins at 250 °C. Between 250 °C and 500 °C, carbide formation increases. Above 550 °C, a decrease in carbide content is observed, with an almost complete disappearance at 800 °C. For the Mid-T baking above 500 °C, the mechanism responsible of the disappearance of carbides during the heating phase has not been fully identified but may be due to the desorption of CO or $CO_2$ (see Fig. 4). Similar results were reported in [7]. To confirm these results, the surface of the Nb samples was also studied using ToF-SIMS. Depth profiling with ToF-SIMS was employed to examine the distribution of carbides beneath the surface and to compare with the data presented in Fig. 5. The ToF-SIMS analysis was carried out at Western University in Ontario (see Fig. 7).

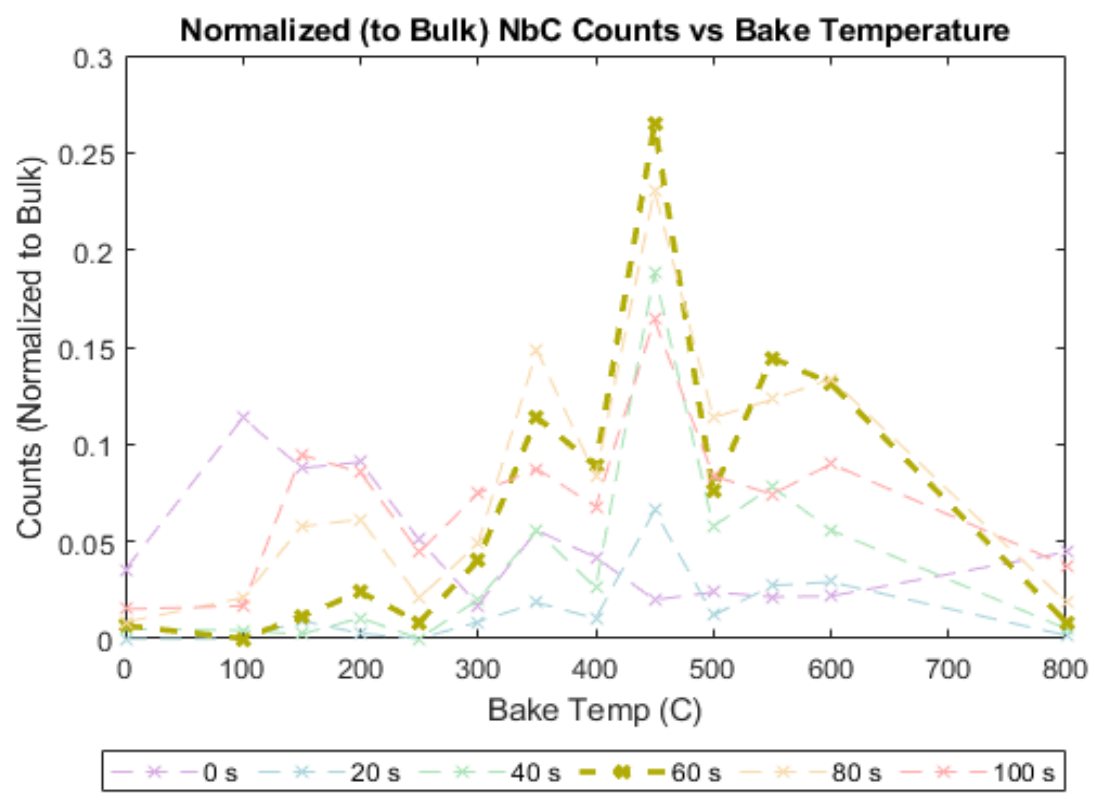

Figure 7: Evolution of niobium carbides from 25 °C to 800 °C as analyzed by XPS.

Figure 7 shows the same trend as Fig. 5. The evolution of carbide matches the XPS profile but with a shift of 50 °C; the maximum is observed at 450 °C. To confirm the absence of surface contamination, the samples were additionally characterized using scanning electron microscopy (SEM, Zeiss). Nb samples were analyzed at each temperature (Fig. 8). The baseline (room temperature) shows no carbide formation on the surface. At 350 °C, the initial stages of carbide formation are observed. It is only at 500 °C that some precipitation (triangular-shaped features) becomes visible on the surface.

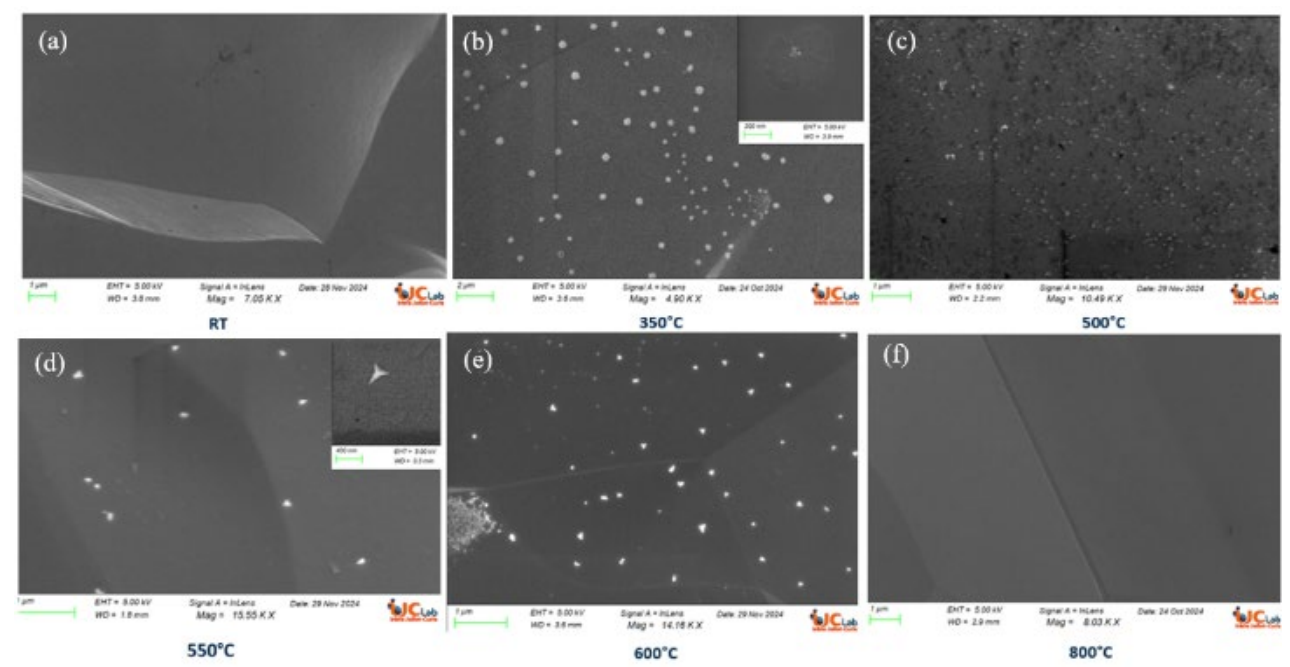

Figure 8: SEM image of niobium without annealing (a), after annealing at 350 °C (b), at 500 °C (c), at 550 °C (d), at 600 °C (e) and 800 °C (f).

Some carbide precipitation is visible at 550 °C, and similarly at 600 °C, until the carbides disappear from the surface at 800 °C. At this temperature, a clean surface is observed, with no evidence of carbide formation or contamination.

## SUPRATECH FURNACE

In this section, we present the results of heat treatments performed using a different furnace at two conditions: 800 °C for 3 hours at $10^{-5}$ mbar, and 600 °C for 10 hours at $10^{-5}$ mbar. The 800 °C heat treatment was carried out to compare the effect of using two different furnaces on identical samples (same background, temperature, and duration). The main difference between the two setups was the vacuum level. The objective of this comparison is to determine whether the quality of the ultra-high vacuum (UHV) environment influences carbide formation. SEM inspection of the niobium samples thermally treated at 800 °C reveals significant differences between the results obtained using the Supratech furnace and the TRIUMF furnace. The SEM image in Fig. 9(a), shows submicron-sized, pyramid-shaped precipitates. Additionally, a heat treatment was carried out at 600 °C for 10 hours. The SEM image at this temperature shows interestingly no evidence of carbide formation (see Fig. 9(b)). These results don't follow the trend observed on samples treated at TRIUMF. The longer duration (10 hrs) at 600 °C may explain this behaviour but additional data are required.

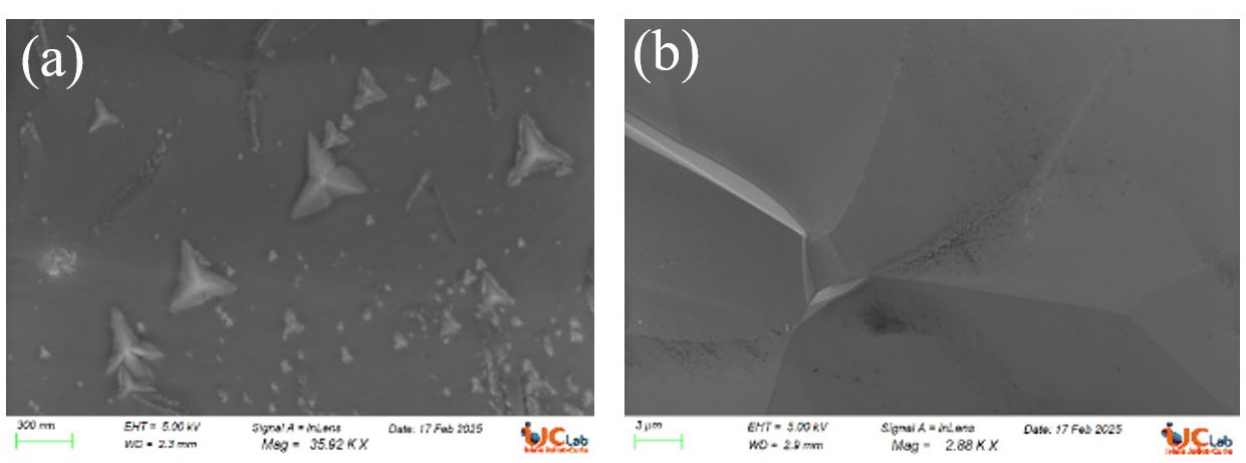

Figure 9 (a): SEM image of niobium annealed at 800 °C for 3 hours under an ultra-high vacuum of $10^{-5}$ mbar. 9 (b): SEM image of niobium annealed at 600 °C for 10 hours under an ultra-high vacuum of $10^{-5}$ mbar.

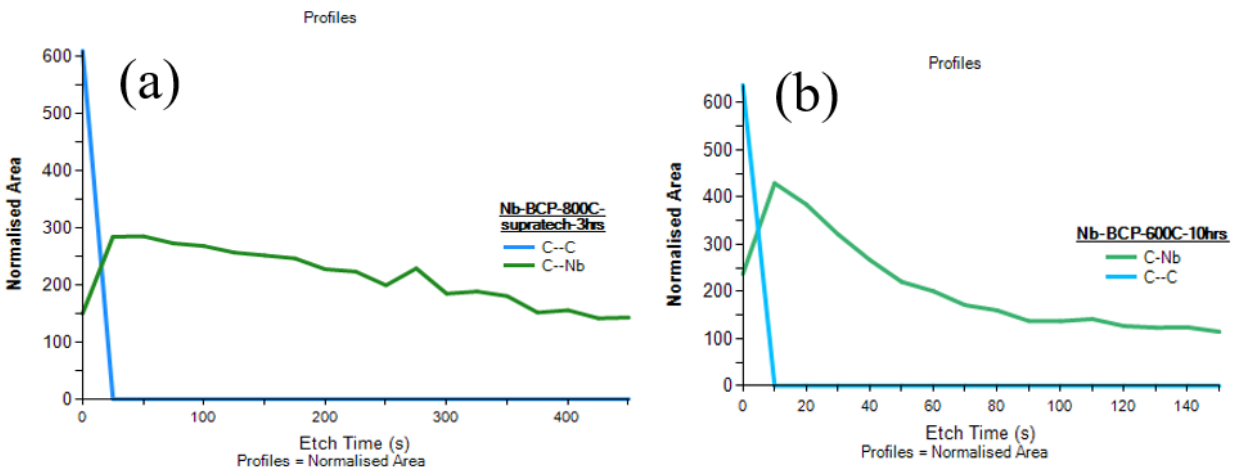

Figure 10: XPS profile of niobium carbide annealed at 800 °C for 3 hours under an ultra-high vacuum of $10^{-5}$ mbar (a). XPS profile of niobium carbide annealed at 600 °C for 10 hours under an ultra-high vacuum of $10^{-5}$ mbar (b).

The carbide profiles (Fig. 10) were obtained through XPS analyses after abrasion (200 eV) for annealing at 800 °C and 600 °C. The profile at 800 °C reveals that carbides are abundant at the surface but continue to extend deeper into the material. At 600 °C, the Nb–C bonding is more pronounced in the first nanometers but decreases faster than at 800 °C. Several 352 MHz Spoke cavities have been heat treated in Supratech furnace at 650 °C during 10 h without any subsequent flash BCP showing very high $Q_0$ above 4E10 at 2 K (see Fig. 11). The sample study results tend to show that degradation of $Q_0$ may be linked to the formation of carbon precipitates at the surface and not necessarily to the presence of Nb-C bonds underneath the surface. Confirming this statement would require additional tests as the samples didn't follow exactly the same surface preparation as a cavity. The samples were indeed not high pressure rinsed with ultra-pure water in a clean room after heat treatments.

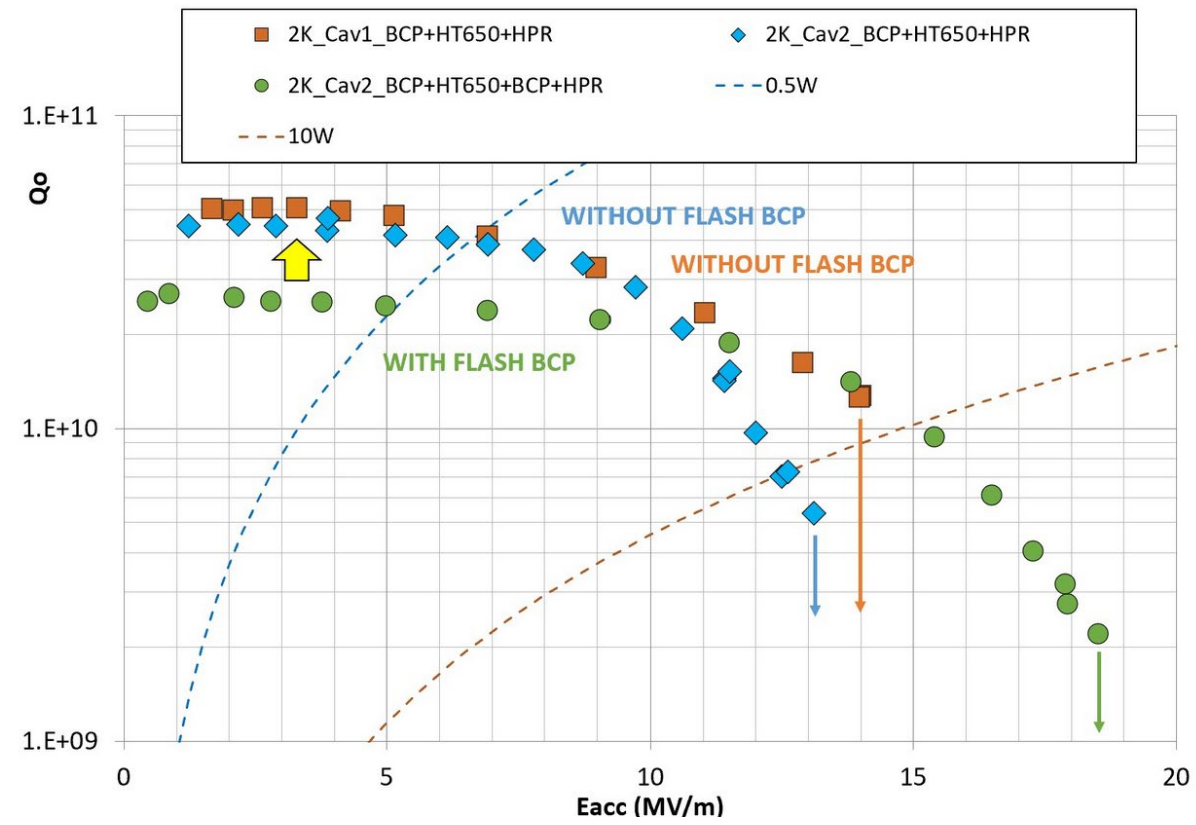

Figure 11: Comparison of $Q_0$-E behaviour measured at 2 K for cavities that were furnace-baked at 650 °C during 10 hrs.

## CONCLUSION

XPS and ToF-SIMS analyses show that niobium carbide formation begins near 250 °C, increases toward ~450–500 °C, then decreases between ~550–700 °C and vanishes at 800 °C under TRIUMF's UHV conditions. Residual-gas analysis shows higher $CO_2$ partial pressure at higher temperature, could explain the disappearance of carbon for heat treatments above 500 °C. SEM observations corroborate these trends: carbide is formed above 350 °C, but structured precipitation appears at 500 °C (triangular/pyramidal precipitates) and disappear at 800 °C. A direct comparison between furnaces reveals the critical role of vacuum quality: in the Supratech furnace (worse vacuum, ~$10^{-5}$ mbar) carbides were abundant at the surface and extended deeper even after an 800 °C bake, whereas in the TRIUMF UHV bake carbides were removed at 800 °C. Notably, some cavities baked at 650 °C during 10 h still exhibited very high $Q_0$. The sample study has to be continued to address at which extent the type of furnace can influence the carbide formation and, as well, if HPR could have a role.